# Classification of superconductors on $T_C$ Map

Wei Fan

*Key Laboratory of Materials Physics, Institute of Solid State Physics, Chinese Academy of Sciences, 230031 Hefei, People's Republic of China*

**Abstract:**

The $T_C$ map in parameter space of electron-phonon interaction $\lambda$ and phonon frequency $\Omega_P$ is used to classify already known superconductors. The $T_C$ map is partitioned into six regions based on superconducting parameters $\lambda$ and $\Omega_P$. We briefly discuss the properties of superconductors in every region. The $T_C$ map can be used as a useful tool to find and design new superconductors with higher $T_C$ and better properties for their real applications.

Email: fan@theory.issp.ac.cn

**Introduction**

Superconductors have several pronounced advantages in electric power transmission over conventional conductors such as copper and aluminum: they can carry much higher currents, most importantly, without energy loss due to zero resistance. The task for finding superconductors with higher transition temperatures is crucial to their real applications. Since the discovery of first superconductor Hg at 4K, the current record of transition temperature $T_C$ has reached about 160K of copper-oxides superconductors. The standard explanation of superconductivity is based on electron-pair theory, which has been proved by the quantization of fluxoid and the effect of Coulomb-blockage of mesoscopic superconducting island [1]. The basic theory that can explain superconductivity is the BCS theory and its strong-coupling generalization [2-3]. In this article we introduce $T_C$ map that we had obtained in a former work [4], and classify some already known superconductors on the $T_C$ map. The parameter space $\lambda$-$\Omega_P$-$\mu^*$ for the $T_C$ map is constructed by $\lambda$ characterizing electron-phonon interaction, $\Omega_P$ characterized phonon frequency and $\mu^*$ measured the Coulomb interaction between electrons. The phonon frequency $\Omega_P$ can be measured from Raman spectrum, infrared spectrum and neutron-diffraction spectrum. The Coulomb pseudo-potential $\mu^*$ is not only hard to measure in experiments but also difficult to calculate theoretically. The renormalized Coulomb pseudo-potential $\mu^*$ are generally equal to 0.1~0.2. The parameter $\lambda=-\partial\Sigma/\partial\omega|_{\omega=0}$ represents the renormalized effects from electron-phonon interaction such as the effective mass of electron and density of states near Fermi energy with the renormalization from electron-phonon interaction are expressed as $m^*=m(1+\lambda)$ and $N(0)^*=N(0)(1+\lambda)$ respectively. Experimentally $\lambda$ can be extracted from phonon line-width in neutron scattering measurements [5] and estimated from parameter $\lambda_{tr}$ in electric-transport measurements [6]. Theoretically, the parameter $\lambda$ can be obtained by the calculation of Eliashberg function $\alpha^2F(\omega)$ using linear response method within the frame of density functional theory [7].

**$T_C$ map on $\lambda$-$\Omega_P$ plane**

The $T_C$ map is plotted with contour lines in Fig.1, which is calculated numerically by using the standard strong-coupling theory with the Coulomb pseudo-potential $\mu^* =0.1$. Two thick blue contour lines with $T_C \approx 30K$ and $T_C \approx 160K$ are the highest $T_C$ before the discovery of cuprate high-temperature superconductor [8] and the current highest $T_C$ of cuprate superconductor [9] respectively. The horizontal line $\lambda=2.0$ is McMillan's upper-bound, for $\lambda>2.0$ superconductivity is unstable because of the structural instability [10]. In fact, $\lambda>2$ is equivalent to $J^2>E_0\Omega_0$, where J is the coupling constant of electron-phonon interaction, $E_0$ the typical energy of electron and $\Omega_0$ the typical energy or frequency for phonon-modes that contribute to superconductivity. We can see that above condition is easily satisfied for low phonon frequency. The soft-mode effects generally increase $\lambda$ and simultaneously induce the structural instability. Therefore, $\lambda<2$ or $J^2<E_0\Omega_0$ sets a limit for superconducting parameters. We can partition the $T_C$ map into six regions labeled from *A* to *F*. There are five regions from *A-E* under the horizontal line $\lambda=2$, one region *F* with $\lambda>2$ has the structural instability.

The superconducting parameters $\lambda$, $\Omega_P$ and $\mu^*$ of superconductors can be obtained from theoretical calculations and experimental measurements. We will introduce the different parameter regions and find superconductors that have superconducting parameters within the corresponding regions. The classification of superconductors on the $T_C$ map is very important to understand already known superconductors and to find the route designing new superconductors with higher $T_C$ and engineering superconductors more suitable to applications.

Before introducing these different regions on the $T_C$ map, we introduce an important relation $<\omega^2>\lambda=\eta/M$ for superconductors, where $\eta$ is the Hopfield parameter which is approximately considered as a constant for a certain class of superconductors, M is effective mass for a phonon mode that has contribution to superconductivity. If we ignore isotope effect, above relation is equivalent to $<\omega^2>\lambda=\eta/M=$constant. The constant measures the possible maximum of transition temperature by $T_C^{Max} \propto (<\omega^2>\lambda)^{1/2}$ for all superconductors with superconducting parameters satisfying $\eta/M$ equal to a fixed value. Now we have two constraints for superconducting parameters : $\lambda<2$ and $<\omega^2>\lambda=\eta/M=$constant.

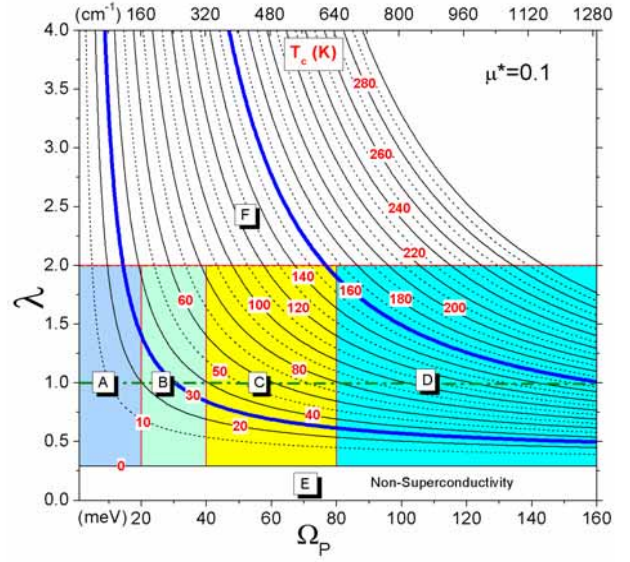

Figure 1. The $T_C$ map on $\lambda$-$\Omega_P$ plane. The labels from *A* to *F* illustrate six different regions partitioned on the map.

*A*: ( 0meV < $\Omega_P$ <20meV ) Heavy metals and their alloys: The first discovered superconductor Hg belongs to this region. The most prominent point in this region is that the lowest phonon frequency and smaller values $<\omega^2>\lambda$. The superconductivity in this region is driven by strong electron-phonon interaction. The parameters $\lambda$ for some of superconductors such as Pb are larger than 1.0 [3,7]. The smaller values of $<\omega^2>\lambda$ decide the possible maximum of $T_C$ is only 30K from the $T_C$ map. Other superconductors in this region are La, Tl, $In_2Bi$, $Sb_2Tl_7$, $Nb_3Sn$ and many others.

*B*: ( 20meV < $\Omega_P$ < 40meV ). In this region the frequencies or energies of phonon increase because of smaller masses of atoms. The superconductors include the light metals and their alloys such as Al, Ti, V, $V_3Si$ and Mo. Some of non-superconducting light-metal such as Li and Ca become superconductors under high-pressure [11]. Their superconductivities come from the enhanced $\lambda$ with increasing pressure but the frequencies have no significant increase with pressure. The recently new founded iron-based superconductors $LaO_{1-x}F_xFeAs$ [12] and $SmO_{1-x}F_xFeAs$ [13] and the intermetallic high-temperature superconductors such as $Nb_3Ge$ are in this region. Although the $Nb_3Ge$ is in region *B*, however it is close to the boundary between *A* and *B*. The superconductivity in this region is driven by strong electron-phonon interaction.

*C*: ( 40meV < $\Omega_P$ < 80meV ) Most of cuprate superconductors with high transition temperature belong to this region, for examples, $YBa_2Cu_3O_7$, $Bi_2Sr_2CaCu_2O_{8+\delta}$ and $HgBa_2Ca_2Cu_3O_{8+\delta}$ with current $T_C$-record 160K under pressure[9]. Both strong electron-phonon interaction with $\lambda>1$ and high phonon frequency guarantee their high transition temperatures. There are strong anisotropic superconductivity, complex crystal structures and chemical composition. The small coherent length and large energy gap make the higher critical magnetic field and in-homogenous superconductivity. The well known non-cuprate high-temperature superconductor $Ba_{1-x}K_xBiO_3$ with $T_C$=30K [14-15] locates in this region as well.

Most interestingly, many recently discovered novel superconductors with components of light elements C and B belong to this region, for example, $MgB_2$ [16-17], Borocarbide superconductors such as $YPd_2B_2C$ [18-19], $Rb_3C60$ and other alkali-fulleride superconductors [20]. Although $\Omega_P$ <80meV, the highest phonon frequency can be larger than 100meV. The $T_C$ is benefit from the high phonon frequency or vibrations of light atoms although the parameters $\lambda$ electron-phonon interaction are small ($\lambda<1$) compared with cuprate superconductors.

*D*: ( 80 meV < $\Omega_P$ < 160meV ). The hydrogen-rich compound such as $SiH_4$ becomes metal and superconductor at high pressure [21]. Theoretical calculations show that other hydrogen-rich compounds $SnH_4$ and $GeH_4$ are hopefully become superconductors at high pressure. The superconductor $SiH_4$ at high pressure is the realization of high temperature superconductor based on the idea of metal hydrogen. The high temperature superconductors under ambient pressure with composition of light atoms such as Be, B, C and N are more desirable for applications. For a room-temperature superconductor, its superconducting parameters should be in this region, so the material should have strong electron-phonon interaction and simultaneously keep the structural stability.

*E*: ( $\lambda$ < 0.30 ) There are non-superconductivity in this region at normal conditions. However at specific condition, some of them can become superconductors such as superconducting alkali-metal Cs at high pressure. Nobel metals such as Au, Ag and Cu aren't superconductors even at high pressure.

*F*: ( $\lambda$ > 2.0 ) Strong electron-phonon interaction can induce large lattice deformation and other structural phase transitions. Some of them, for instance charge-transfer insulating cuprates, become superconductors after having introduced carriers by doping the parent materials, or changing lattice space by introducing small molecule such as $H_2O$ into the parent materials, for examples, $Na_xCoO2 \cdot yH_2O$ [22] and $SrFe_2As_2 \cdot yH_2O$ [23]. However for some superconductors such as pyrochlore $KOs_2O_6$, $T_C$ decreases by introducing $H_2O$ molecules [24].

Generally, there are no clear boundaries between different regions on the $T_C$ map. The Zr nitridochlorides superconductors $Li_xZrNCl$ are located at the boundary between regions *B* and *C* [25-26]. The superconductor $SiH_4$ at high pressure and organic superconductors such as κ-phase BEDT-TTF salts are belong to both *C* and *D* because they have very high energy phonon up to 180 meV but the effective phonon energy $<\omega>_{ln}$ are generally smaller than 80 meV.

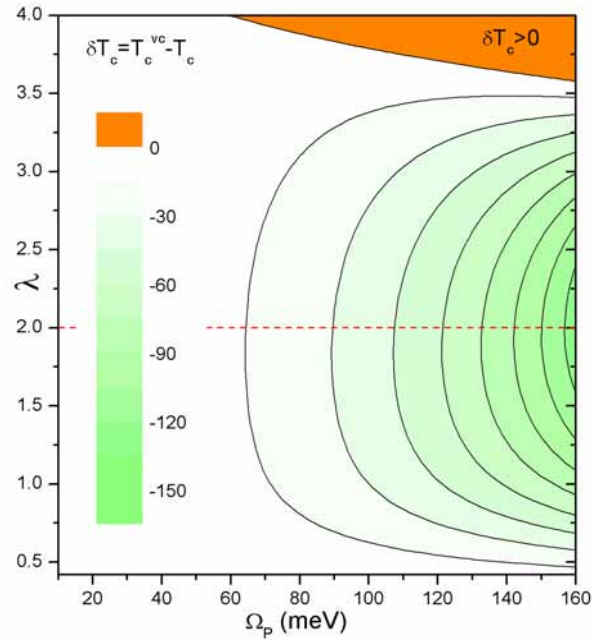

Figure 2. $T_C^{VC}$ is the $T_C$ including vertex correction and $\delta T_C$ is the change of $T_C$. The control parameter of vertex correction is $E_b$=2eV (band width or Fermi energy).

**Some discussions**

The superconductivity with high $T_C$ is closely relative to non-adiabatic effects. Theoretically, we should consider the vertex correction [4]. The effect of vertex correction on $T_C$ map is plotted in Fig.2. We can see that the vertex corrections depress $T_C$ significantly at high frequency near horizontal line $\lambda \approx 2.0$ (the McMillan upper limit). It's clearly that in region *D* we should consider the effects of vertex corrections. If the half band width $E_b$ is smaller than 2eV, vertex correction will introduce more serious problems. So we must estimate $T_C$ based on the Fig.1 and Fig.2 simultaneously. Experimentally, the structural instability under high strain is the bottleneck for the realization of high-temperature superconductors in *D* region.

The magnetism of material is ignored, so for superconductors with the coexistence of superconductivity and magnetism, the $T_C$ map overestimates $T_C$ value. Additionally, the electron-pairing mechanism for copper oxides high-temperature superconductors has no decisive answer. However, this doesn't prevent their classifications on the $T_C$ map. In fact, the evidences of electron-phonon interaction as the pairing mechanism have been accumulated enough to explain the superconductivity of copper-oxides high-temperature superconductors (see a review [27]).

For materials with small sizes such as metal films on insulating substrates or small particles with nano-sizes, additional ingredients are needed to explain their superconductivity. In a bulk material, there are very dense electronic energy levels and the mean level space $\delta E$ is far less than superconducting energy gap $\Delta_{SC}$ or $\delta E << \Delta_{SC}$. However, in a nano-particles, the mean level space $\delta E$ increases with decreasing size, so below a critical size $\delta E > \Delta_{SC}$, the superconductivity will be destroyed by effects of small sizes [28]. Additionally, strong quantum fluctuation and phase slip also depress the superconducting state of superconducting film and contribute the residual resistance below $T_C$ [29]. The critical sizes for superconducting films are generally 1-10 nanometers dependent on materials.

If the sizes of confined dimensions are much larger than above critical size, the superconductivities of nano-structures are very similar to their bulk materials. However there are still interesting things. The granular Bismuth nano-wires are superconductors at ambient pressure however their bulk materials aren't superconductors [30]. The $T_C$ of granular Bismuth nano-wire is close to $T_C$ of bulk Bismuth at high pressure. Another example metal Platinum isn't superconductor however its powder samples are superconductors [31]. The starting pressure for superconducting Boron nano-wires is significantly reduced to smaller values compared with its bulk materials [32]. There are intrinsic lattice deformations near gain-boundaries so the enhanced $T_C$ in nano-material is probably induced by the larger strains and stronger electron-phonon interactions in grain-boundaries. Thus it's very hopefully to find superconducting nano-materials with higher $T_C$.

**Conclusion**

In this article, we introduce the $T_C$ map and its non-adiabatic correction. The $T_C$ map is used to classify some already known superconductors based on their phonon frequencies and parameters of electron-phonon interaction. The magnetism and finite size effects of superconductors are also briefly discussed. The $T_C$ map is very helpful to guide to design new superconductor with higher $T_C$ and better properties for real applications.

**Acknowledgment**
This work is supported by Director Grants of Hefei Institutes of Physical Sciences, Knowledge Innovation Program of Chinese Academy of Sciences and National Science Foundation of China.